\documentclass{elsarticle}

\usepackage[utf8]{inputenc} 
\usepackage[T1]{fontenc}    
\usepackage{hyperref}       
\usepackage{url}            
\usepackage{placeins}
\usepackage{booktabs}       
\usepackage{amsfonts}       
\usepackage{nicefrac}       
\usepackage{microtype}      
\usepackage{lipsum}
\usepackage{graphicx}
\usepackage{graphics}
\usepackage{xcolor}
\usepackage{subcaption}
\usepackage{overpic}
\usepackage{natbib}
\usepackage[margin=1.5in]{geometry}

\setcitestyle{authoryear,open={(},close={)}}

\begin{document}

\begin{frontmatter}
\title{Using Social Networks to Improve Group Transition Prediction in Professional Sports}
\author[byu1]{Emily J. Evans}
\author[byu1]{Rebecca Jones}
\author[byu1]{Joseph Leung}
\author[byu1]{Benjamin Z. Webb}

\address[byu1]{Department of Mathematics,
  Brigham Young University,
  Provo, Utah 84602, USA}


\begin{abstract}

{We examine whether social data can be used to predict how members of Major League Baseball (MLB) and members of the National Basketball Association (NBA) transition between teams during their career. We find that incorporating social data into various machine learning algorithms substantially improves the algorithms' ability to correctly determine these transitions. In particular, we measure how player performance, team fitness, and social data individually and collectively contribute to predicting these transitions. Incorporating individual performance and team fitness both improve the predictive accuracy of our algorithms. However, this improvement is dwarfed by the improvement seen when we include social data suggesting that social relationships have a comparatively large effect on player transitions in both MLB and in the NBA.}
\end{abstract}

\begin{keyword}
sports, social networks, machine learning, group transition\\

{\it MSC2010:} 91D30, 62P25, 91B40 
\end{keyword}

\end{frontmatter}

\section{Introduction}
Social connections exist between and across many different types of groups. These could be social relations between individuals in different schools, families, religious and professional groups, or any other group defined by affiliation. The importance of these social connections is that they are not always between members of the same group. 

In this paper, we address how these relations affect individual transitions between groups. The social connections we study are those formed between members of professional sports teams, specifically the social connections between the teams in the MLB (Major League Baseball) and the teams in the NBA (National Basketball Association). The transitions we study are the player transitions from one team to another within MLB and the NBA, respectively, during a player's professional career.

These baseball and basketball teams can be thought of as specific types of \textit{professional groups}, i.e., a group of individuals employed by the same employer, with a similar skill set, and with a specific objective. Transitions between such professional groups, although similar in some ways, are not the same as transitions between \textit{social communities}, which are communities defined in terms of social interactions~\citep{Girvan7821}.  The dynamics of individual, or more generally, node transition between social communities is a relatively new field taking cues from mathematics~\citep{CommTransition1} and computer science~\citep{CommTransitionBayesian}.  As this is not the focus of this paper, we refer the interested reader to a survey of work~\citep{CommTransitionSurvey}.

Transitions between professional groups have been studied by sociologists and economomists, see for example ~\citep{StrengthWeakTies, Coworkernetwork, labourmobility, Marsden2001, productiveemployee, CALVOARMENGOL2005500, referal, Motivate, inequality, SNESLMO}.  To understand these transitions features such as the strength of ties between workers~\citep{StrengthWeakTies}, the geography of transition~\citep{labourmobility}, the role social networks play in finding employers and employees~\citep{Coworkernetwork, Marsden2001} are considered. In regards to the social network aspect of such transitions, specific questions that have been considered are whether companies hire workers through social ties of productive employees~\citep{productiveemployee, SNESLMO}, whether employees hired through referrals are more likely to stay~\citep{referal}, how unemployed workers find jobs through their social networks~\citep{CALVOARMENGOL2005500, inequality}, differences in salaries between employees found by referrals~\citep{Coworkernetwork}, and motivations for employees to grow their professional social networks~\citep{Motivate}.     

Since social connections often exist between individuals in different professional groups, a natural question is how these connections influence transitions between these groups. Here we consider the specific question of how social data compared to other player and team statistics can be used to improve our ability to predict the way professional athletes transition from one professional group to another, i.e., one team to another in the MLB and NBA.

Other ways of studying group transitions that are specific to these teams include analyzing the labor market's influence on professional baseball and basketball players, which has been studied extensively beginning with the classic work of Rottenberg~\citep{baselabourclassic}. The transition between teams in baseball has been studied in light of changes to rules governing transitions~\citep{baserevisit} and also in terms of player productivity~\citep{baselaborallstar}. In professional basketball, hiring decisions have been considered in light of first-hand experience~\citep{basketwhohire} and also in terms of increased productivity~\citep{BasketballProduct}. Moreover, tools from network theory have been used to study interactions in both baseball~\citep{SAAVEDRA20101131} and basketball~\citep{Fewell2012BasketballTA, Clemente2015NetworkAI}.  However none of the listed works consider how a player's social network influences transitions between teams despite the fact that social network analysis has become increasingly popular in sports analytics~\citep{Wasche2017138}.

From the various professional groups that exist, the reason we choose to analyze the team dynamics of the MLB and NBA is the availability of data. This includes the player's social data but also information such as the player's performance, and other factors that could be used to predict transitions between teams. The size of the data set, measured in terms of the number of individuals, the number of years it spans, and variety of statistics is also important as our analysis relies on machine learning algorithms that require sufficient amounts of data to decrease bias and improve accuracy (see Section~\ref{sec:ML} and~\citep{Murphy}).

To address the question of what influences transitions between professional teams we consider three factors: individual performance, team fitness, and social data. Of the three, \textit{individual performance} is perhaps the most natural to consider. The reason being is that poor performance presumably motivates managers to remove players while high performance makes players more attractive to other teams~\citep{baselaborallstar,basketwhohire, BasketballProduct}. See also Sections~\ref{sec:baseper} and~\ref{sec:baskper}.

While individual performance is important in understanding transitions between teams, alone it gives us relatively little information about ``where'' a player will transition to (see Sections \ref{sec:resultsumm} and \ref{sec:4}). To understand a professional athlete's tendency to move from one team to another we use team \textit{fitness} together with \textit{individual performance} in our analysis. The idea is that an athlete with high performance is more likely to transition to a team he or she perceives as either being fit, or becoming more fit~\citep{martens1971group}. An athlete with low performance is potentially more likely to get traded to a team with lower fitness.  

In the larger context of group dynamics there are many ways to measure \textit{fitness} including how cohesive or stable the group is~\citep{stadtfeld2020emergence}, the strength of individual members, and the ability of the group to perform it designated task. In this study, we considered two proxies which we use to measure the fitness of our groups. The first is \textit{relative team ranking}, which acts as a measure of a team's ability to achieve success. The second proxy for team fitness is the \textit{financial valuation} of a team, which is based on the notion that a team on firm financial footing is more stable and can likely offer high performers more competitive salaries~\citep{willyerd2014high}.      

The third factor we consider is the \textit{social interactions} individuals have within the network of professional groups (see Section~\ref{sec:basesocial}). If the player has social connections to other players from other teams, then this may indicate at least a predisposition to move to that team. The majority of the data we use in our study to determine whether two players are interacting is Twitter data. Specifically, we use Twitter data to create two multilayered social networks with nodes representing individual players and directed edges indicating who follows who, one for the MLB and one for the NBA. These networks act as a proxy for the social interactions among the players (see Figure~\ref{fig:basenetworkdata}) of both sports. In addition to Twitter data, for the NBA we also use the college a player attended to create a proxy social network.  The idea is that players may feel a connection with other alumni from their schools who were trained in a similar way by the same coach. As the data on college attendance for the MLB is largely incomplete we only create this proxy for the NBA.

The overarching question is how the combination of these three factors-- \textit{individual performance}, \textit{team fitness}, and \textit{social interaction}-- allows us to predict to which team an individual will transition to.  More specifically, we are interested in whether having a knowledge of social connections will increase the accuracy of our predictions.

What we find is that the addition of performance data, team fitness data, and social data each improve the predictive ability of the machine learning algorithms we consider for both the MLB and the NBA. Performance and team fitness, perhaps surprisingly, only modestly raise the accuracy of these algorithms. The inclusion of social data from Twitter, however, dramatically improves the predictive ability of these algorithms in every case we consider. Additionally, including the college a player attended, which is our second proxy for social data in the NBA, similarly increases the performance of these algorithms but not as much as Twitter data. (See Section \ref{sec:resultsumm} for a summary of these results.)   

Over time an increasing number of players in both the MLB and NBA begin to follow other players. An interesting feature of the Twitter data we have is that the fraction of players that do so in the NBA is significantly higher than in the MLB (see Figure \ref{fig:Twittercompare}). It is possible that this explains to some extent why our predictions are better for the NBA than the MLB when using Twitter data (see Section \ref{sec:resultsumm}). To add evidence to this, when we limit our predictions to the latter decade of our study when Twitter use is at its highest, we can predict transitions much more accurately for both MLB and the NBA than if we try to predict transitions for the first decade (see Tables \ref{tab:baseballsplit} and \ref{tab:basketsplit}).

We also find that although the Twitter networks for baseball and basketball are fairly different in size the two networks are strikingly similar. Specifically, they have very similar network statistics including mean degree, fraction of nodes in the largest strongly connected component, mean distance between connected node pairs, clustering coefficient, reciprocity, and the degree assortativity (see Table \ref{tab:basicstats}). Therefore, it seems unlikely that the particular structure of these networks can explain why Twitter data leads to higher accuracy for basketball when compared to baseball.

The paper is organized as follows. In Section \ref{sec:resultsumm} we give a brief summary of our results regarding prediction accuracy in both MLB and the NBA. In Section \ref{sec:meth} we describe our methodology including which social and nonsocial data we collected and some of the features of this data. This includes performance, fitness, social, and other data we used to train the machine learning algorithms we selected. In Section \ref{sec:ML} we give a brief description of these algorithms. In Section \ref{sec:4} we describe how different combinations of social and nonsocial data improve the accuracy of these algorithms. In Section \ref{sec:netanal} we analyze the basic statistics of the baseball and basketball Twitter networks. We conclude in Section \ref{sec:con} with some open questions that specifically relate to how this type of analysis could be extended to study group transitions in other settings, i.e., other professional groups and more general multilayered social networks.

\section{Summary of Results}\label{sec:resultsumm}
Here we give a brief summary of the results found in Section \ref{sec:4} regarding the accuracy of the machine learning algorithms we consider. The different types of data we use to determine player transition between teams are broadly speaking \textit{player performance}, \textit{team fitness}, and \textit{social data}, which are described in detail in Section \ref{sec:meth}. 

We find that the addition of each of these data-types each improve the predictive ability of our algorithms over the years we consider. As mentioned, performance data by itself, however, does little to raise the accuracy of these algorithms. Specifically, including performance data raises the accuracy of these algorithms by at most $1.5\%$ for the MLB and $0.15\%$ for the NBA over the probability of $1/29\approx3.45\%$ of a correct random guess. Similarly, using team fitness data improves accuracy by at most $0.5\%$ for the MLB and at most by $1.5\%$ for the NBA. Using all nonsocial data together including performance, team fitness, player position, team, and career length improves accuracy by at most $1.7\%$ for the MLB and $5.1\%$ for the NBA (see Tables \ref{tab:basesummary} and \ref{tab:basketsummary}).   

When using social data the situation improves dramatically. When using data derived from Twitter connections, with no other information, the prediction accuracy of the algorithms can be as high as $16.9\%$ for the MLB and $26.1\%$ for the NBA, an increase of $13.4\%$ and $22.6\%$ over random guessing, respectively. Using college data for the NBA similarly increases the accuracy of prediction to as much as $11.2\%$ and using both Twitter and college data together increases accuracy up to $26.7\%$. Combining all social and nonsocial data brings our highest performing algorithms up to an accuracy of $19.4\%$ for the MLB and $29.7\%$ for the NBA for the years we consider. It is worth noting that our maximum accuracy is found in the MLB using only the player's team together with Twitter data while in the NBA our maximum accuracy is found using only the team's fitness combined with Twitter data (see Tables \ref{tab:basesummary} and \ref{tab:basketsummary}).  

As mentioned in the introduction, over time an increasing number of players in both the MLB and NBA have begun to follow other players (see Figure \ref{fig:Twittercompare}). When we limit our predictions to the last decade of our study when Twitter use is at its highest, we can predict with up to $21.1\%$ of the time where a player will transition to in the MLB and up to $33.1\%$ of the time in the NBA (see Tables \ref{tab:baseballsplit} and \ref{tab:basketsplit}). This is our highest accuracy for any of the cases we consider, suggesting that the more players use Twitter, the easier it is to predict to what team they with transition to in the following season. 

\section{Methodology}\label{sec:meth}

\subsection{Baseball performance dataset}\label{sec:baseper}
The first professional sport we consider is professional baseball.  Major League Baseball consists of 30 teams evenly split between the American and National league.  Each full team roster consists of 40 players. A baseball season consists of 162 regular season nine inning games with some positions playing most games, and some positions like pitcher playing in a fraction of these games. In our analysis we consider 3 high-level positions that players can be in: pitcher, catcher, and fielder, where the position of \textit{fielder} represents all other positions. We note that positions are more fine grained, but typically players who play in the infield and outfield have some flexibility in the actual position that they play.  We singled out the catcher position because, usually, one of the catchers serves as team captain. It is worth noting that the exact composition of a team's roster varies with some teams having more of one position than another.  

The baseball performance data for the 2002 -- 2018 seasons we use were obtained from \texttt{https://www.baseball-reference.com}.    
Although the website contains a wide variety of statistics such as number of games played, points scored, and total hits for our analysis we focused primarily on a few advanced statistics and a few engineered statistics instead of generic totals.  The data collected for a player includes the main position played, the team played on, and the player's age for a given season. 
In addition we collected the following advanced data for each player: the \textit{fielding percentage} (FLD\%), \textit{offensive winning percentage} (OWn\%), \textit{adjusted batting runs} (BtRuns), and \textit{adjusted batting wins} (BtWins). 

OWn\% is the percentage of games that a team would win if the batting was done by $9$ copies of the player, assuming average offense and defense. BtWins estimates a player's total contribution to his team's wins with his bats.
BtRuns is an estimate of a player's running contribution to a team's wins. FLD is the number of putouts and assists divided by the sum of putouts, assists, and fielding errors. This data provides an overall picture of a player's performance during the season.
While other metrics are often used in evaluating player performance, we selected metrics that were representative of both pitching and catching positions and were available on \texttt{https://www.baseball-reference.com}.




We then created the following engineered data for each player and each season:  

\begin{itemize}
\item Position -- created by merging actual players positions into the three positions we identified: pitcher, catcher, and fielder.
\item Career length -- number of prior seasons played until the year under consideration (i.e., rookies have a career length of zero).
\item Leave variable -- specifies if a player is to leave their current team after the season under consideration. 
\item Target variable -- specifies which team a player plays for the next year, or if they do not return to play that next year. 
\end{itemize}
The \textit{leave variable} is critical in identifying which players transition at the end of the season to another team allowing us to focus on predicting the transitions of those players. The \textit{target variable} provides us the ground truth for measuring the accuracy of our results.  

To illustrate our collected and engineered features we display three seasons of data for Giancarlo Stanton in Table~\ref{tab:gscareer}.  We note that at the end of 2017, Giancarlo switched teams, (to the New York Yankees), hence the engineered field of target was set to NYY.

\begin{table}[h]
    \centering
    \begin{tabular}{ccccccccc}
    \hline
    \hline
         Season & Position & Team & FLD\% & OWn\% & BtWins & BtRuns & CL & Target\\
         \hline
         2016 & FD & MIA & .982 & 1.2 & .585 & 11.9 & 7 & N/A \\
         2017 & FD & MIA & .998 & .735 & 59.8 & 5.6 & 8 & NYY \\
         2018 & FD & NYY & .992 & .621 & 26.7 & 2.6 & 9 & N/A \\
    \hline
    \hline
    \end{tabular}
    \caption{Three years of collected and feature engineered data for Giancarlo Stanton.  We observe that at the conclusion of the 2017 season, Giancarlo transitioned from the Miami Marlins (MIA) to the New York Yankees (NYY). Thus in 2017 his target value is set to NYY. In this table, and in Table~\ref{tab:dccareer1} we use the abbreviation CL for career length (an engineered variable).}
        \label{tab:gscareer}
\end{table}

We show the specific distribution of players, players leaving their team, players retiring, and players transitioning for each year in Table~\ref{tab:baseball_year_data}. 
We note that each year approximately 50\% of players leave their team each year in some manner.

\begin{table}[h]
    \centering
    \begin{tabular}{ccccccccc}
    \hline
    \hline
    Season & 2002 & 2003 & 2004 & 2005 & 2006 & 2007 & 2008 & 2009 \\
    \hline
    \# of Players & 1090 & 1124 & 1125 & 1127 & 1116 & 1162& 1166 & 1142 \\
    Total Leaving & 527 & 606 & 560 & 605 & 513 & 566 & 565 & 531\\
    Retiring & 236 & 274 & 276 & 283 & 257 & 302 & 288 & 281\\
    Switched Teams& 291 & 332 & 284 & 322 & 256 & 254 & 277 & 250\\
    \hline
    \hline
    \end{tabular}
    \vspace{.2in}
    
    \begin{tabular}{cccccccccc}
    \hline
    \hline
    Season & 2010 & 2011 & 2012 & 2013 & 2014 & 2015 & 2016 & 2017 & 2018\\
    \hline
    \# of Players & 1151& 1160 & 1176 & 1194 & 1204 & 1243 & 1233 &  1222 & 1266\\
    Total Leaving&  538 & 551 & 565 & 562 & 610 & 633 & 585 &578 & 601 \\
    Retiring& 287&271&273&281 &279 & 321 & 326 & 288 & 285\\ 
    Switched Teams& 251& 280 & 292& 281& 331& 312& 259& 290&316\\
    \hline
    \hline
    \end{tabular}
    \caption{The number of Major League baseball players per year in our data set comprising the 2002-2018 baseball seasons.  We observe that each year 48.7\% of the players leave their current team on average.  Of those that transition about one-half, 50.3\%, of the players transition to a new team and the other half of the players end their professional careers.}
    \label{tab:baseball_year_data}
\end{table}



\subsection{Basketball performance dataset}\label{sec:baskper}

The second professional sport we considered is professional basketball.  Similar to baseball, the National Basketball Association consists of 30 teams evenly split between two conferences.  In the NBA, each team's roster consists of only 17 players, with only eight players required to be active at any one time.  Basketball has five positions: point guard, shooting guard, small forward, power forward, and center; however most basketball players are capable of playing in more than one of the positions.  Each team plays 82 games in a standard season.  

The basketball performance data for the $2001$--$2018$ seasons was collected from \url{https://www.basketball-reference.com}. Similar to baseball, in our analysis, we choose to use advanced data statistics, focusing on three advanced stats. 
PER, \textit{Player Efficiency Rating}, measures how much a player produced in one minute of play. 
\textit{Win Shares} or WS is an estimate of how many wins were contributed by a player.
PBM, \textit{Box Plus/Minus}, is an estimate of the number of points per $100$ possessions that a player contributed.  To illustrate both the collected and engineered statistics we consider a few seasons of DeMarcus Cousins' career in Table~\ref{tab:dccareer1}, and note that he switched teams in 2018.

\begin{table}[h]
    \centering
    \begin{tabular}{ccccccccc}
    \hline
    \hline
         Season &  Team & PER & WS & BPM & CL & Target\\
         \hline
    2017 & NOP & 23.2 & 1.6 & 5.5 & 6 & N/A\\       2018 & NOP & 22.6 & 4.7 & 4.7 & 7 & GSW\\  
    2019 & GSW & 21.4 & 2.4 & 3 & 8 & N/A\\
    \hline
    \hline
   \end{tabular}
    \caption{Three years of collected and engineered data for DeMarcus Cousins.  At the conclusion of the 2018 season, DeMarcus transitioned from the New Orleans Pelicans (NOP) to the Golden State Warriors (GSW) so the target variable was set to be GSW.}
    \label{tab:dccareer1}
\end{table}

Similar to baseball, we also created engineered features for individuals each season. Since basketball has only 5 positions, we did not modify this feature, and only engineered values for career length, leave and target. Ultimately, there were $1614$ basketball players who switched teams between $2001$ and $2018$.  The distribution of the leaving players is shown in Table~\ref{tab:my_label}.  The average percent of players leaving their team each year is $60\%$, and approximately $33\%$ of those that leave retire.

\begin{table}[h]
    \centering
    \begin{tabular}{cccccccccc}
    \hline
    \hline
    Season & 2001 & 2002 & 2003 & 2004 & 2005 & 2006 & 2007 & 2008 & 2009 \\
    \hline
    \# of Players & 441 & 440 & 428 & 442 & 464 & 458 & 458 & 450 & 444 \\
    Total Leaving& 251 & 238 & 284 & 289 & 282 & 245 & 253 & 258 & 259\\
    Retiring& 84 & 43 & 85 & 78 & 62 & 98 & 89 & 85 & 79\\
    Switched Teams& 167 & 153 & 206 & 227 & 184 & 156 & 166 & 173 & 180 \\
    \hline
    \hline
    \end{tabular}
    \vspace{.2in}
    
    \begin{tabular}{cccccccccc}
    \hline
    \hline
    Season & 2010 & 2011 & 2012 & 2013 & 2014 & 2015 & 2016 & 2017 & 2018\\
    \hline
    \# of Players & 442 & 452 & 478 & 468 & 481 & 492 & 476 & 486 & 540\\
    Total Leaving&  281 & 257 & 295 & 273 & 267 & 279 & 259 & 270 & 310\\
    Retiring &  71 & 73 & 100 & 83 & 86 & 96 & 90 & 94 & 128\\ 
    Switched Teams& 210 & 184 & 195 & 190 & 181 & 183 & 169 & 176 & 182\\
    \hline
    \hline
    \end{tabular}
    \caption{The number of National Basketball Association (NBA) players per year in our data set comprising the 2001-2018 seasons.  We note that there are approximately half the number of players in this dataset compared to the baseball dataset. Also, a higher percentage of players transition annually in the NBA (on average about 60\%). However the ratio of retiring players to players switching teams is approximately the same as in baseball. }
    \label{tab:my_label}
\end{table}

\begin{figure}
\begin{center}
\includegraphics[width=.6\linewidth]{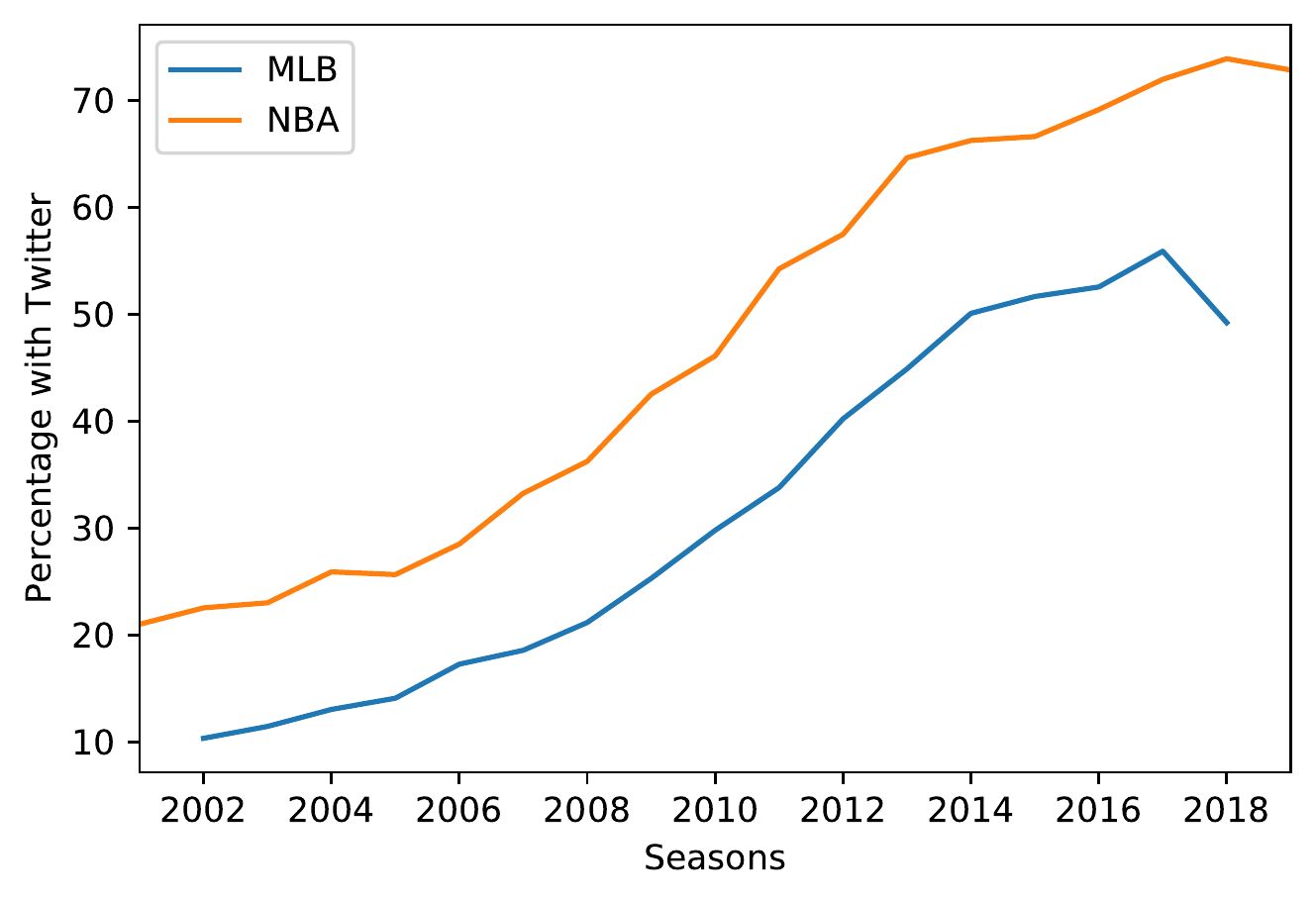}  
\caption{A comparison of the percentage of players who had an active Twitter account before and including July 2020 and who played between 2002 -- 2018 for the MLB (blue) and between 2001 -- 2019 for the NBA (orange).  The datasets we use only give the number of players that have Twitter and who played in a given year not the specific players that used Twitter during that year.  This explains why there is Twitter data as early as 2001, even though Twitter began in 2006.}
\label{fig:Twittercompare}
\end{center}
\end{figure}

\subsection{Social network datasets}\label{sec:basesocial} 

Beyond performance data for each player, we conjectured that a player's social network would impact team transition.  Acknowledging that it would be extremely difficult to impossible to create a ground truth social network for players, we created an approximation of a social network utilizing Twitter data.  Twitter is a social networking site that allows users to exchange 140 character ``Tweets'' with followers.  Twitter was chosen because player Twitter handles were available from  from both \url{https://www.basketball-reference.com} and \url{https://www.baseball-reference.com}, and because Twitter provides an easy API that allowed us to obtain both the followers and those followed by a user.  One potential downside of using Twitter is that the ``followers'' information is not time stamped.  Hence the social network created with the Twitter data is a snap shot of the relationships that existed before and up to July 2020 when we scraped the data with no way to pinpoint when a player started to follow another.\\

\begin{figure}[h]
\begin{center}
    \begin{overpic}[scale=0.35]{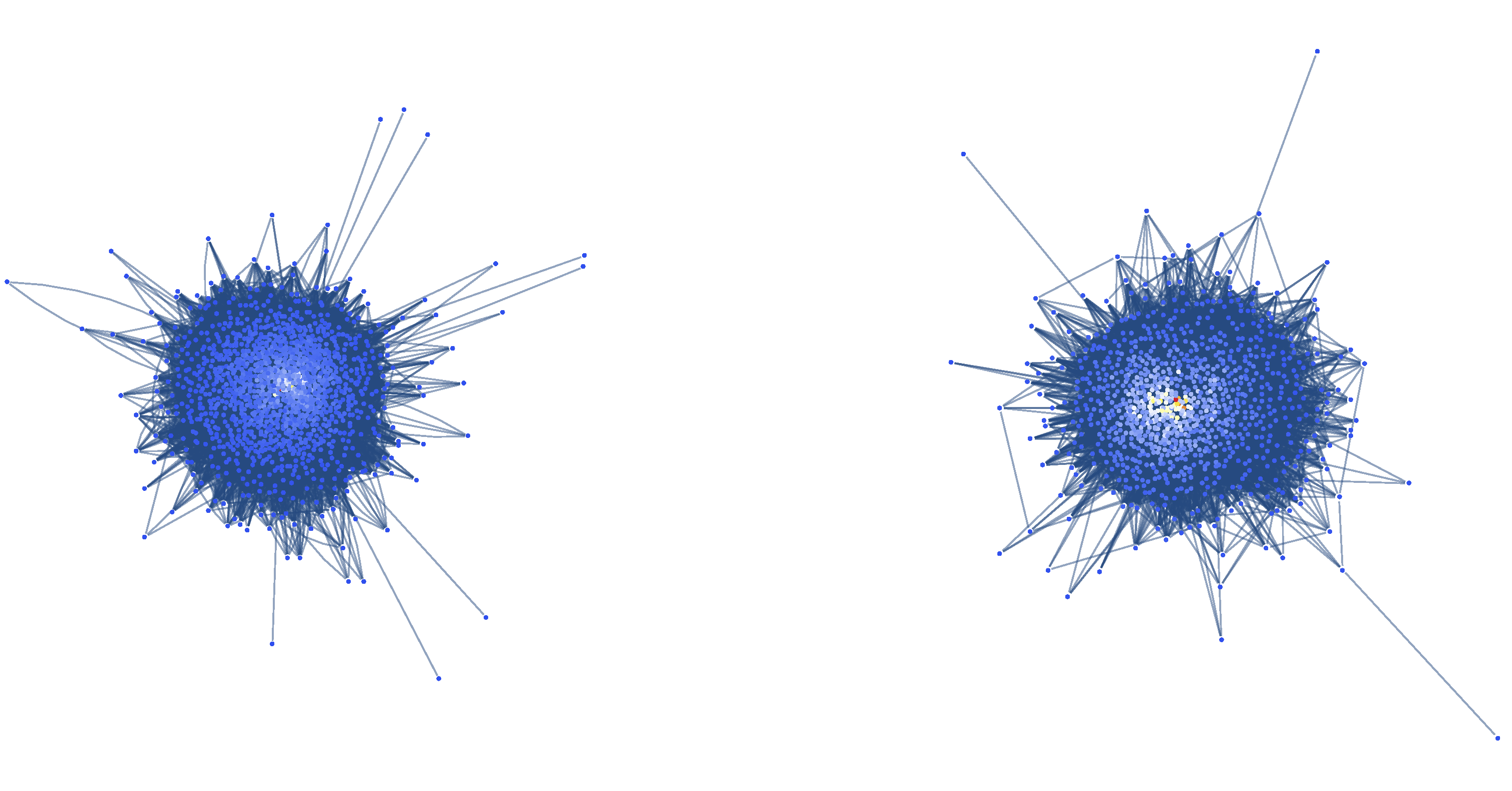}
    \put(4,2){baseball Twitter network}
    \put(63,2){basketball Twitter network}
    \end{overpic}
    \caption{The baseball Twitter network and the basketball Twitter network are shown left and right, respectively. Players are colored by number of neighbors, i.e. \textit{degree centrality} equal to the number of followers $+$ number followed. Brighter colors indicate more neighbors.}
    \label{fig:basenetworkdata}
\end{center}
\end{figure}

\begin{figure}
    \begin{center}
    \begin{tabular}{cc}
        \begin{overpic}[scale=0.4]{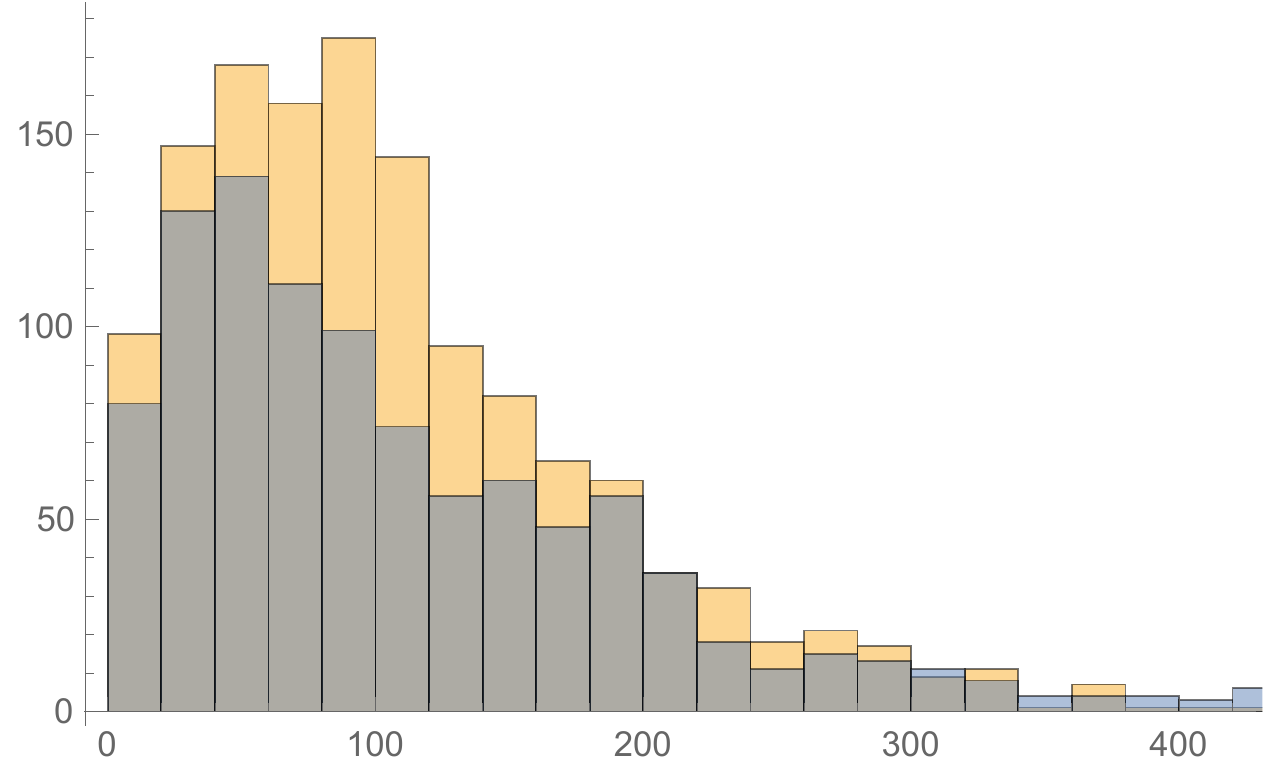}
        \put(2,-7){MLG and NBA degree centrality}
        \end{overpic} & \hspace{0.5cm}
        \begin{overpic}[scale=0.4]{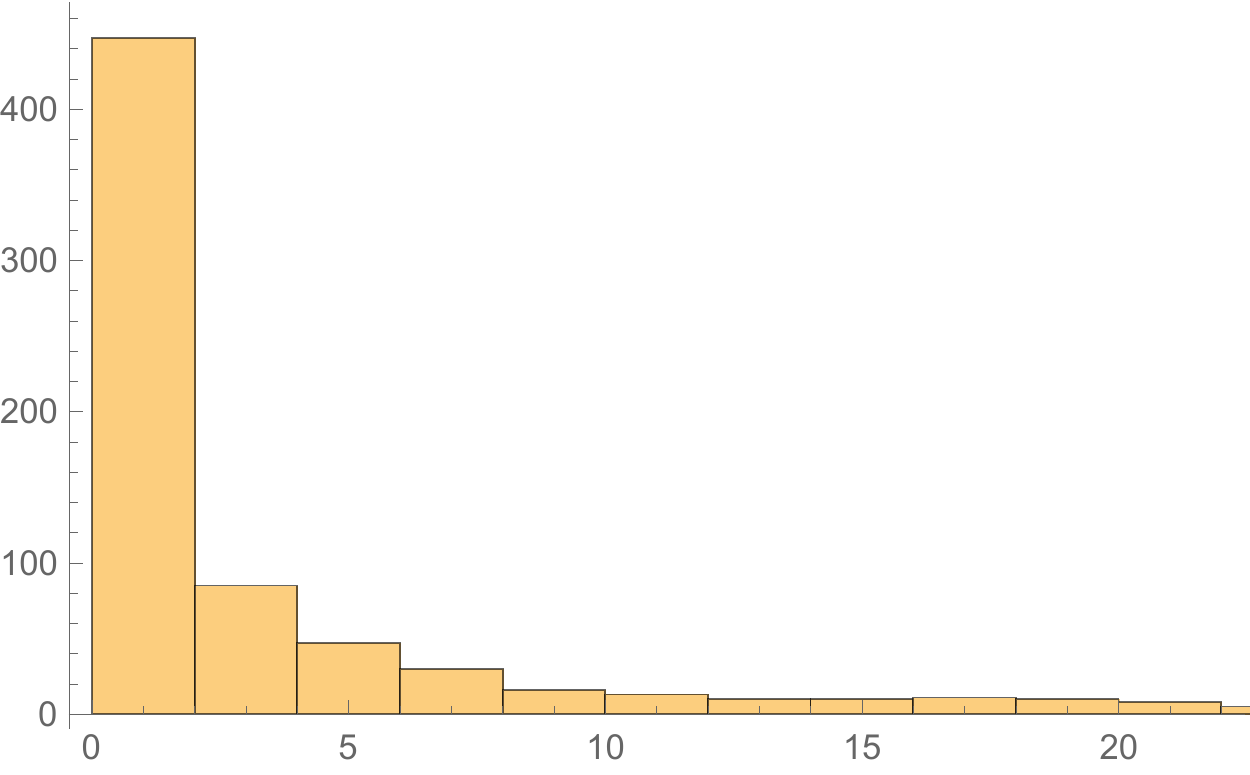}
        \put(20,-7){NBA college attendance}
        \end{overpic}
    \end{tabular}
    \vspace{0.2cm}
    \caption{Left: A histogram of the degree centrality for the baseball Twitter network and the basketball Twitter network shown in orange and grey, respectively. Right: A histogram of the number of colleges that had one of more basketball players join the NBA from 2001--2019.}
    \label{fig:degreecentrality}\label{fig:college_attendence}
    \end{center}
\end{figure}

With the Twitter data we created a directed social network of players where player A has a connection directed to Player B if Player A followed Player B, which we refer to as our baseball Twitter network and basketball Twitter network, respectively. Of the $4207$ unique baseball players that switched to a different team from 2002 -- 2018, we were able to obtain Twitter handles for $702$ of them.  For basketball players that switched between 2001 -- 2019, we were able to collect Twitter handles for $784$ of the $1847$ players, a significantly larger percentage (see Figure \ref{fig:Twittercompare}). The resulting Twitter network is a social network with $53690$ directed edges for baseball and $43827$ directed edges for basketball (see Figure~\ref{fig:basenetworkdata}).  Most players in both datasets have a relatively small number of connections or \textit{degree centrality} to others, which is the number of followers together with number of players followed for a specific player. A few players do have a large number of connections though (over 100).  The distribution of connections for both baseball and basketball players having at least one Twitter connection is shown in Figure~\ref{fig:degreecentrality} (left). (A more thorough analysis of these networks is given in Section \ref{sec:netanal}.) 

For the NBA we also investigate whether the college a player attended can serve as a proxy for social connections and whether this data helps predict where player's transfer during their professional career. To test this idea, we pulled the college data for each basketball player from \url{https://www.basketball-reference.com} and created a categorical feature for colleges. If a player in this set did not go to college, they were included and their college category was \textit{N/A}. 

There were $259$ different colleges attended by future NBA players in the data set. Figure~\ref{fig:college_attendence} (right) shows the number of colleges in the data set that had a given number of players attend. The \textit{N/A} category is excluded. For example, from the data there were $5$ schools that each had between $15$ and $17$ players attend: Syracuse: 16, Michigan State: 15, Georgetown: 17, LSU: 15, and Georgia Tech: 16. $940$ players did not attend a college.

\begin{figure}
\begin{center}
    \begin{overpic}[scale=0.33]{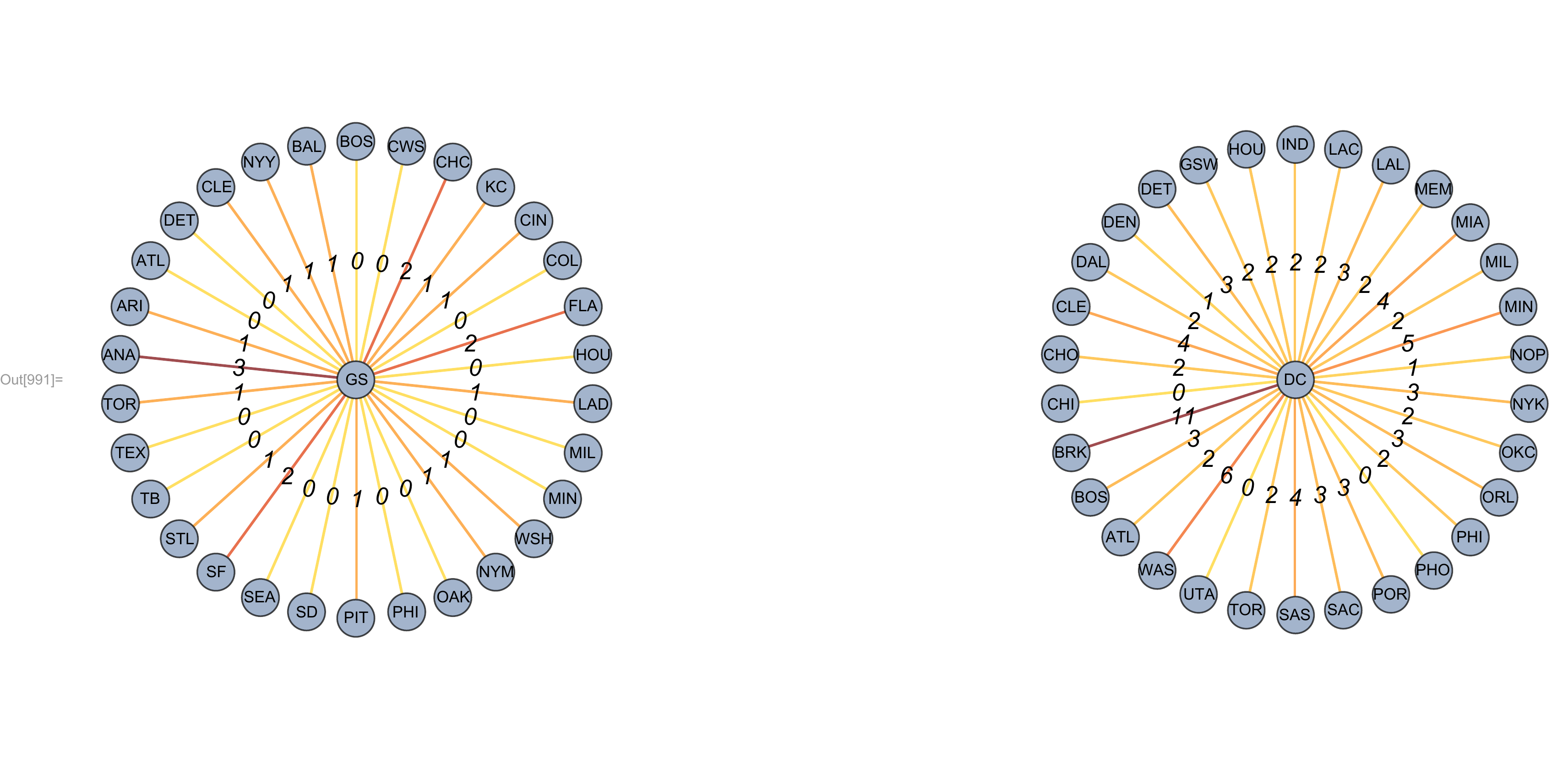}
    \put(3,2){Giancarlo Stanton (MLB)}
    \put(67,2){DeMarcus Cousin (NBA)}
    \end{overpic}
    \caption{Left: The affinity network of the MLB player Giancarlo Stanton (GS) from the 2017 season. Right: The affinity network of the NBA player DeMarcus Cousin (DC) from the 2018 season. In both, edge weights indicate the player's affinity score where darker edges indicate a higher score.}
    \label{fig:affinity}
\end{center}
\end{figure}

Using our Twitter networks and the team each player played on for a given season we create an \textit{affinity network} for each player as follows: For a given transitioning player we add a weighted edge connecting the player to each of the teams in the MLB/NBA. The weight of an edge is the number of other players from that team this player \textit{followed} during that season, which we call the affinity score (see, for example, Figure \ref{fig:affinity}).  We emphasize that the social network between players is fixed across seasons but the social affinity score \textit{changes} between seasons since players change teams. This weight gives a score of the affinity that a player has for the team for a given season. Since we do not allow a player who has been identified as \textit{transitioning} to remain on their current team we set the affinity score for the current team to zero. Finally the way we handle mid-year transitions (i.e., midyear trades) is different between the two sports.  In basketball, for the purpose of affinity score we consider only the team the player was on at the beginning of the season.  For baseball, due to the way information is presented at \texttt{baseball-reference.com} we omit players who transitioned during mid-season from the calculation of the affinity score for a given year.

\subsection{Team stratification engineered data}
Conjecturing that the prestige of a team may influence where a player will move, we also collected two metrics to measure team fitness. First we guessed that teams with more funds, typically resulting in higher pay and more highly paid players, would be more desirable.  We collected data on each team's dollar valuation for each year in question through Forbes.com. We also retrieved team rankings for each year from \url{https://www.basketball-reference.com} and \url{https://www.baseball-reference.com}
since we conjectured that players might be more interested in moving to successful teams. Then we added these features to our dataset based on a player's current team for a season.

Extending Table~\ref{tab:dccareer1}, we have the following additional columns given in Table~\ref{tab:DCcareer}.

\begin{table}[h]
    \centering
    \begin{tabular}{ccccc}
    \hline
    \hline
         Season & Team & Rank & Value & Target\\
         \hline
    2017 & NOP & 20 & 750 & N/A\\       
    2018 & NOP & 8 & 1000 & GSW\\  
    2019 & GSW & 2 & 3500 & N/A\\
    \hline
    \hline
   \end{tabular}
    \caption{Three years of collected team data for DeMarcus Cousins.}
    \label{tab:DCcareer}
\end{table}

\section{Analysis Techniques}\label{sec:ML}
In this section, we describe the techniques used to \textit{make} the group transition predictions.  We utilize a wide variety of machine learning methods, although most can be classified as ensemble methods, to predict to which team a player will transition.  Ensemble methods combine several models (predictors) which operate independently and are typically very good for classification problems. The outputs of the models are then combined into a single prediction that is often better than any single model.  We use four types of ensemble methods to make predictions which can be classified into two different categories: (i) \textit{Randomized decision trees} which include (a) Random Forests and (b) Extremely Randomized Trees, and (ii) \textit{boosting algorithms} which include (c) Adaptive Boosting and (d) Extreme Gradient Boosting.  In addition to ensemble methods, we also use a number of (iii) \textit{other algorithms} including (e) Logistic Regression and (f) $k$-Nearest Neighbors as these are also commonly used for classification.  

\textbf{Random Decision Trees:} The idea behind randomized decision trees is rooted in the construction of a classification tree.  A classification tree takes input data, moves through the various decisions nodes of the tree to a leaf, and returns as output the most common result in that leaf. At each level of a classification tree a decision node is constructed by considering the features not already split on, choosing the best feature to split on, and then choosing the optimal split point~\citep{Murphy}. Random decision trees randomize the creation of classification trees in two ways.  The first is that a random subset of the data is sampled and from that subset a classification tree is created.  A standard classification tree considers all of the remaining features when deciding which feature to split on. However, this often results in trees that are highly correlated.  Instead of using all the remaining features at each level of the decision tree, a random decision tree also chooses a random subset of the remaining features to split on.  This offers two advantages, the first is a collection of uncorrelated trees, and the second is splitting on fewer features results in faster algorithms. 

The random forests algorithm (Forest) operates in essence by having many decision trees, which are trained on different random parts of the training set, ``vote'' on the final classification~\citep{breiman1999random}.  A typical random forest consists of thousands of these voting decision trees, and it is typically the case that some of them are actually good models.  Random forests rely on Condorcet's Jury Theorem from political science which guarantees a collection of weak voters will arrive at the correct decision with high probability~\citep{condorcet_2014}.

In the extremely randomized trees (ExtraTrees) algorithm, instead of looking for an optimal splitting threshold for each feature at each step of the decision tree, thresholds are created at random.  The splitting rule for each tree is chosen to be the best of these random thresholds.  This results in trees that are created more quickly and lower variance in the model at the cost of a slight increase in bias.  These trees similarly ``vote'' as in the case of the random forests algorithm.

\textbf{Boosting Algorithms:} Boosting algorithms are a family of algorithms whose aim is to create a strong learner from a weak learner. Boosting algorithms work by applying the weak learner sequentially to weighted versions of the data where in each sequential application misclassified data is given additional weight.  The weak learner can be any classification or regression model, but the most frequently used learner is a decision tree~\citep{Murphy}.  Important to the construction of a boosted tree is the choice of a loss function, which measures the predictive error of the model.  The goal of boosting is to minimize this loss function, and this is done sequentially using the idea of gradient descent.  

In our work we consider two different boosting algorithms that use decision trees as learners: Adaptive Boosting (AdaBoost) and eXtreme Gradient Boosting (XGBoost).  Adaboost~\citep{adaboost} was the original boosting algorithm and has the characteristic that the decision trees have a single split, sometimes called decision stumps. XGBoost~\citep{chen2016xgboost} is a more recent algorithm for boosting and combines decision trees with more splits and sophisticated algorithms to improve the time it takes for the algorithm to converge to the optimal tree. XGBoost is extremely popular due to its ease in configuring, its relative speed in running and its high accuracy. 

\textbf{Other Algorithms:} In addition to ensemble methods we also use two other classification methods.   The first is multinomial logistic regression (softmax regression) which uses a combination of the softmax function and the technique of regression to construct a multi-class classifier~\citep{Murphy}.  In the context of our problem, multinomial Logistic Regression (LogReg) returns a probability vector where each entry in the vector is a probability that the player transitions to a given team.  The second classification technique we use is that of $k$-nearest neighbors (KNN)~\citep{knn}.  In this technique $k$ nearest neighbors are chosen and a probability of a player transitioning to a given team is the proportion of those neighbors that belong to the given team.

\section{Results}\label{sec:4}
Before presenting the results, we recall the overarching question ``Does a player's social network influence which team the player transitions to?''  As described in the previous section we will apply a variety of machine learning techniques with and without social network information as a feature to answer this question.  We note that in both sports the number of teams is 30, however once we have identified a given player as transitioning to a new team we prohibit the player from transitioning to their current team.  Hence each transitioning player has 29 possible teams to transition to, and the na\"ive probability of transitioning to a given team is approximately $3.45\%$.
\begin{table}[h]
    \centering
    \begin{tabular}{cccccccccc}
    \hline
    \hline
    Season & 2001 & 2002 & 2003 & 2004 & 2005 & 2006 & 2007 & 2008 & 2009 \\
    \hline
    MLB & NA & 23 & 31 & 30 & 48 & 43 & 39 & 62 & 61 \\
    NBA& 41 & 31 & 48 & 60 & 50 & 43 & 58 & 72 & 82\\
    \hline
    \hline
    \end{tabular}
    \vspace{.2in}
    
    \begin{tabular}{cccccccccc}
    \hline
    \hline
    Season & 2010 & 2011 & 2012 & 2013 & 2014 & 2015 & 2016 & 2017 & 2018\\
    \hline
    MLB & 69 & 94 & 119 & 123 & 151 & 148 & 138 & 157 & 169\\
    NBA&  123 & 121 & 127 & 141 & 150 & 145 & 128 & 150 & 148\\
    \hline
    \hline
    \end{tabular}
    \caption{The number of MLB and NBA players in the final dataset, which contains those players who switched to a different team and had a Twitter account. }
    \label{tab:final_numbers}
\end{table}

We consider only those players for which we have social data.  The total number of players for each year is shown in Table~\ref{tab:final_numbers}.  We used a 70/30 random train test split over these years and calculate the accuracy of each model in the test set. Here, the accuracy is the number of teams correctly predicted divided by the total number of teams. Each algorithm was run $10$ times, and the final accuracy shown here is the mean of the accuracy over these $10$ runs.

We analyze the outcome of using different combinations of features with different machine learning algorithms. Although we observe some minor variation with using non-social features, the largest variation occurs when social data, i.e. the Twitter affinity score described in Section \ref{sec:basesocial}, is either included or excluded. The details of these results are described in the following two subsections. 

\begin{table}
\centering
\begin{tabular}{c  c  c  c}
\hline
\hline
Features      & Social & TopMLA  & Accuracy (\%)\\
\hline
Position      &  No      &  Forest  & 4.955      \\
              & Yes & XGB      & 17.566     \\
\hline
Team ID         &    No    & Forest   & 4.491      \\
               & Yes & XGB      & 19.955     \\
\hline
Career Length &   No      & Forest   & 5.221  \\    
              & Yes & XGB      & 17.278     \\
\hline
Performance   &  No      & KNN      & 4.890      \\
              & Yes & XGB      & 16.615     \\
\hline
Rank \& Value &  No      & LogReg   & 4.004      \\
              & Yes & XGB      & 17.832     \\
\hline
Social  Only       & Yes & XGB      & 16.880     \\
\hline
All Features      &  No      & KNN      & 5.133      \\
              & Yes & XGB      & 19.402     \\
\hline
\hline
\end{tabular}
\vspace{.1cm}
\caption{Baseball Prediction Accuracy: The prediction accuracy is shown for team transition in MLB during the time period 2002-2018 for players who had Twitter accounts. Each row indicates which feature(s) were used. The top performing machine learning algorithm (TopMLA) on this task from those described in Section~\ref{sec:ML} is shown along with the algorithm's accuracy over this time period.}
\label{tab:basesummary}
\end{table}

\subsection{Baseball Results}

We summarize the results of our machine learning experiments for the MLB in Table~\ref{tab:basesummary}. In the table each row indicates which features are used. For instance, in the first row only knowledge of the player's position is used to predict where the player transitions to. In the second row both the player's position and the player's social network, i.e. affinity scores, are used. A more complete summary of the data can be found in Table~\ref{tab:baseball} located in the Appendix. Here, we observe that including social data always has a positive effect that varies from improving the algorithms' accuracy from between 12\% to 15.5\% for the most effective algorithms.  We note that, typically, XGBoost and random forest provided the highest accuracy of the algorithms tried.  Moreover, each of the individual non-social feature sets (positions, career length, performance, and team fitness) yield approximately the same accuracy level, and the combination of all these features did not significantly improve any algorithm's accuracy. This suggests that these individual features are either in some sense \textit{linearly dependent}, i.e. they are imparting the same information about a particular player, or that the features work against each other in some way.

The more interesting result is how the addition of social information effects the algorithms' prediction accuracy.  When only social data is included accuracy jumps from 4-5\% to nearly 17\%.  Moreover, when social data is used in combination with other features accuracy increases as much as 15\% to a maximum of 19.4\%. This is strong evidence that for baseball who you know, or at least ``who you follow'', strongly influences which team you will transition to.

\begin{table*}
\centering
\begin{tabular}{c  c  c  c}
\hline
\hline
Features      & Social & TopMLA  & Accuracy (\%)\\
\hline
Position      &  No      &  LogRes  & 3.723      \\
              & Yes & Forest      & 28.225     \\
\hline
Team          &    No    & XGB   & 6.364      \\
               & Yes & ExtraTrees      & 27.142     \\
\hline
Career Length &   No      & Forest   & 5.802  \\    
              & Yes & Forest      & 27.445     \\
\hline
Performance   &  No      & LogRes      & 3.593     \\
              & Yes & ExtraTrees      & 28.009    \\
\hline
Rank \& Value &  No      & Forest   & 6.032      \\
              & Yes & ExtraTrees      & 30.238     \\
\hline
Twitter Only       & Yes & ExtraTrees      & 26.104     \\
College Only       & No & ExtraTrees      & 11.169    \\
All Social & Yes & ExtraTrees & 26.667\\
\hline
All Features      &  No      & Forest      & 8.571      \\
              & Yes & Forest      & 29.740     \\
\hline
\hline
\end{tabular}
\vspace{.1cm}
\caption{Basketball Prediction Accuracy: The prediction accuracy is shown for team transition in the NBA during the time period 2001-2018 for players who had Twitter accounts. Each row indicates the feature(s) used. The top performing machine learning algorithm (TopMLA) on this task from those described in Section~\ref{sec:ML} is shown along with the algorithm's accuracy. We note that a ``yes'' in the social column implies Twitter data was used. The ``All Social'' row includes both Twitter and College data.}
\label{tab:basketsummary}
\end{table*}

\subsection{Basketball Results}
A summary of the results for the NBA can be found in Table~\ref{tab:basketsummary}, and, as with the results for the MLB, a more complete summary can be found in Table~\ref{tab:basketdata} in the Appendix.  In Table~\ref{tab:basketsummary} we see that adding social data improves performance remarkably with an increase of over 22\% accuracy in every case.  Like baseball, the ``non-social'' features had very little impact on accuracy. Performance data alone is slightly better than random, while using only social data results in a high accuracy. Adding social data improves accuracy across all features, and using only social data is worse than using social data with any other feature. Using team specific information such as rank and valuation with social data achieves the highest accuracy.

We also experimented with predicting where a player transitions using all players who had \textit{college data}, i.e. had attended college. The players who had college data but did not have Twitter data were excluded from our initial data set, so we wanted to know if using this data on the larger set of players who attended college would result in better predictions. 

In all cases we find that using college data increased prediction accuracy (see Table~\ref{tab:collegedata} in the Appendix) but not as much as including social data from those who use Twitter (cf. Table~\ref{tab:basketsummary}). We find that using college data alone results in a max accuracy of $10\%$, which is less than the result on the smaller data set of those with Twitter data (see Tables~\ref{tab:collegedata} and ~\ref{tab:basketsummary}, respectively). Career length with the college data resulted in a maximum accuracy of $16\%$ when compared to any other single feature, i.e., Team, Position, Performance, Rank and Value (see Table~\ref{tab:collegedata}). 

The fact that college data, which we consider to be a form of social data, can increase our prediction accuracy beyond anything besides Twitter data gives strength to the argument that social connections are an important aspect of predicting the ``where'' of team transitions. College data also has the important trait that it was collected before transitions happened as opposed to Twitter data that was compiled after the fact. Hence, there is no ambiguity as to whether a transition influenced the formation of a social connection or whether the social connection influenced the transition.  

\begin{table*}
\centering
\begin{tabular}{c c c c c c c c c}
\hline
\hline
             &  SocD    & PerD  & CareerL   & Position & TeamRV & 2002-       & 2010-  & 2002- \\
             & & & & & & 2009 & 2018 & 2018\\
\hline
1       &  yes           & yes       & yes      & yes   & yes   &   7.451    & 21.082   &   19.402\\
2       &  yes           & yes       & no       & no    & no    &    6.569     & 17.290     &   16.615\\
3       &  yes           & no        & yes      & no    & no    &  7.843    & 18.889   &  17.278 \\
4       &  yes           & no        & no       & yes   & no    &    10.00     &  19.288    &   17.566\\
5       &  yes           & no        & no       & no    & yes   &    8.039    &   18.718   & 17.832 \\
6       &  yes           & no        & no       & no    & no    &     7.941  &  19.829   &   16.880 \\
7       &  no            & yes       & no       & no    & no    &   6.863   &  4.615   &    4.890 \\
8       &  no            & no        & yes      & no    & no    &  7.059    &  5.185   &   5.221 \\
9       &  no            & no        & no       & yes   & no    &   7.549    &  5.470    & 4.955    \\
10      &  no            & no        & no       & no    & yes   &   5.980    &  4.786  &  4.004 \\
\hline
\hline
\end{tabular}
\vspace{.1cm}
\caption{Baseball Prediction Accuracy: The prediction accuracy using Extra Trees for team transition in MLB during the time period 2002-2019 is shown. Each row indicates whether social data (SocD), performance data (PerD), career length (CareerL), position (Position), and Rank/Value (TeamRV) were used in the experiment. }
\label{tab:baseballsplit}
\end{table*}
\begin{table*}
\centering
\begin{tabular}{c c c c c c c c c}
\hline
\hline
             &  SocD    & PerD  & CareerL   & Position & TeamRank & 2001-       & 2010-  & 2001- \\
             & & & & & & 2009 & 2018 & 2018\\
\hline
1       &  yes           & yes       & yes      & yes   & yes   &   21.940    &   33.135   & 29.740    \\
2       &  yes           & yes       & no       & no    & no    &  22.537 &  32.919        & 28.009   \\
3       &  yes           & no        & yes      & no    & no    &  22.388      &  32.297    & 27.445   \\
4       &  yes           & no        & no       & yes   & no    &  22.164      &   32.270  & 28.225   \\
5       &  yes           & no        & no       & no    & yes   &  23.358    &   31.297   & 30.238   \\
6       &  yes           & no        & no       & no    & no    &  21.343      & 32.459     & 26.667   \\
7       &  no            & yes       & no       & no    & no    &  4.179      &  4.676   & 3.507    \\
8       &  no            & no        & yes      & no    & no    &  4.253       &   4.622   & 5.802    \\
9      &  no            & no        & no       & yes   & no    &  4.478       &    4.324   & 3.679    \\
10      &  no            & no        & no       & no    & yes   &   6.866    &  5.919    & 6.032   \\
\hline
\hline
\end{tabular}
\vspace{.1cm}
\caption{Basketball Prediction Accuracy: The prediction accuracy using Extra Trees for team transition in NBA during the time period 2001-2019 is shown. Each row indicates whether social data (SocD), performance data (PerD), career length (CareerL), position (Position), and team rank and valuation (TeamRank) were used in the experiment. }
\label{tab:basketsplit}
\end{table*}

\subsection{A temporal comparison}
We conclude this section with one additional comparison.  Recall that we are using Twitter relationships to estimate the social network of players.  There are at least two weaknesses to this approach.  The first is that relationship data received via the Twitter API, which was collect in July 2020, is not time stamped.  Hence it is impossible to discern whether players became followers before or after a transition was made.  The second weakness of this approach is that Twitter was not founded until 2006, and although players from the early years of our study have joined Twitter a much lower percentage of these players have accounts and thus our proxy social network is less complete for those years (see Figure~\ref{fig:Twittercompare}).  

With this in mind we considered one additional test of the efficacy of using Twitter data by comparing the accuracy of our machine learning algorithms for the earlier years (2001-2009) and the later years (2010-2018). The results are shown in Tables~\ref{tab:baseballsplit} and~\ref{tab:basketsplit} for MLB and the NBA, respectively. In every case considered in these tables, if social data is used the algorithm's accuracy is significantly higher in the later time period when Twitter usage is higher than in the earlier time period (see Figure \ref{fig:Twittercompare}). For baseball, using Twitter data alone increases accuracy from $7.9\%$ to $19.8\%$ as the average Twitter usage climbs from $16.4\%$ to $45.3\%$ during 2002--2009 and 2010--2018, respectively. For basketball, accuracy increases from $21.3\%$ to $32.4\%$ as the average Twitter usage climbs from $28.7\%$ to $64.3\%$ during 2001--2009 and 2010--2019, respectively. This suggests that the more complete our information is on the social interactions of players the better we can predict their transitions.

\section{Network Analysis of the Twitter MLB and NBA Data Sets}\label{sec:netanal}
In this section we investigate the properties of both the MLB Twitter and NBA Twitter networks described in Section \ref{sec:meth} (see Figure \ref{fig:basenetworkdata}). We first consider the basic statistical properties of these networks then compare their degree, eigenvector, closeness, and betweenness centralities.  

\begin{table*}[b]
\centering
\begin{tabular}{c  c  c  c  c  c c c c}
\hline
\hline
Network        &  $n$          & $m$       & $c$        & $S$       & $\ell$ & $C$    & $r$     &$a$\\
\hline
MLB            & 1364     & 76977   & 56.43    &  0.931   & 2.07 & 0.189 & 0.605  & -0.043              \\
NBA            & 1003     & 58750   & 58.57    &  0.971   & 2.15  & 0.190 & 0.614 & -0.023              \\
\hline
\hline
\end{tabular}
\vspace{.1cm}
\caption{Basic statistics for the MLB and NBA Twitter networks using Mathematica. Properties measured are: total number of nodes $n$, number of directed edges $m$; mean degree $c$; fraction of nodes in the largest strongly connected component $S$; mean distance between connected node pairs $\ell$; clustering coefficient $C$; reciprocity $r$, and the degree assortativity $a$.}
\label{tab:basicstats}
\end{table*}

The basic network statistics we consider are the network's total number of nodes $n$, number of directed edges $m$, mean degree $c$, fraction of nodes in the largest strongly connected component $S$, mean distance between connected node pairs $\ell$, clustering coefficient $C$, reciprocity $r$, and the degree assortativity $a$. The \textit{mean degree} of the network is $c=m/n$. A \textit{strongly connected component} of a network is a maximal set of nodes such that it is possible to reach any node from any other node. If the largest of these components has $n_{max}$ nodes then $S=n_{max}/n$. The \textit{distance} $d_{ij}$ from node $i$ to node $j$ is the length of the shortest path from node $i$ to node $j$ through the network. If such a path exists we say node $i$ is \textit{connected} to node $j$. The average $\ell=\langle d_{ij}\rangle$ over all connected nodes is the network's \textit{mean distance} between connected nodes. The \textit{clustering coefficient} $C$ is, roughly speaking, the fraction of triangles in the network versus ``potential triangles'' or paths of length 2. The network \textit{reciprocity} is the percentage of edges that are reciprocated or, for our networks, how often a player follows someone that follows them. Last, if the tendency is for players that follow many players to follow those that also follow many players then the network is said to be \textit{assortative} where $0<a\leq1$. Otherwise, the network is \textit{disassortative} with $-1\leq a<0$. (For a more detailed description of these network quantities see \citep{Newman2010}.)    

In Table \ref{tab:basicstats} these statistics are shown for both networks. Although the number of nodes and edges in these networks are, relatively speaking, quite different each of the other statistics in the table are very similar. In fact, it is striking how similar some of these statistics are. This suggests that these two networks have very similar structures which in turn suggests that the reason we have better predictions for the NBA versus the MLB is not due to specific structural features of the networks.

\begin{figure}
\begin{center}
    \begin{overpic}[scale=0.44]{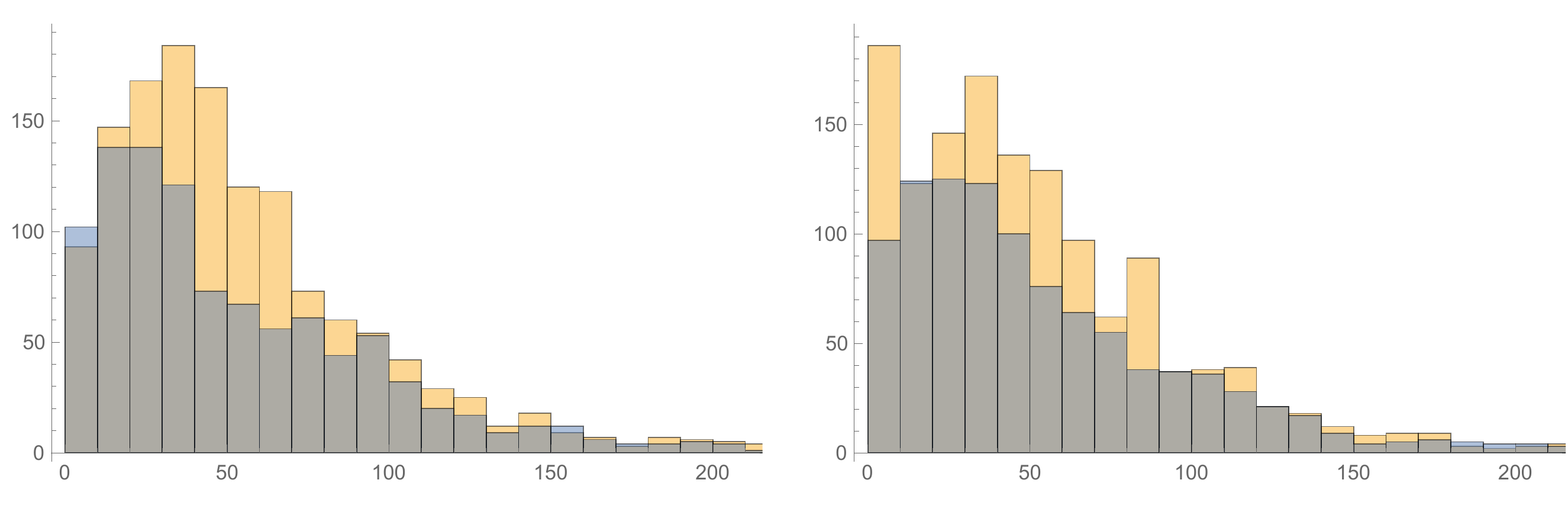}
    \put(13,-2){in-degree centrality}
    \put(63,-2){out-degree centrality}
    \end{overpic}
    \vspace{0.25cm}
    \caption{Histogram of the in-degree and out-degree centrality for the baseball and basketball-Twitter networks shown left and right, respectively. Baseball is shown in orange and basketball is shown in blue in each histogram.}
    \label{fig:degreecentrality1}
\end{center}
\end{figure}

\begin{figure}
\begin{center}
    \begin{overpic}[scale=0.33]{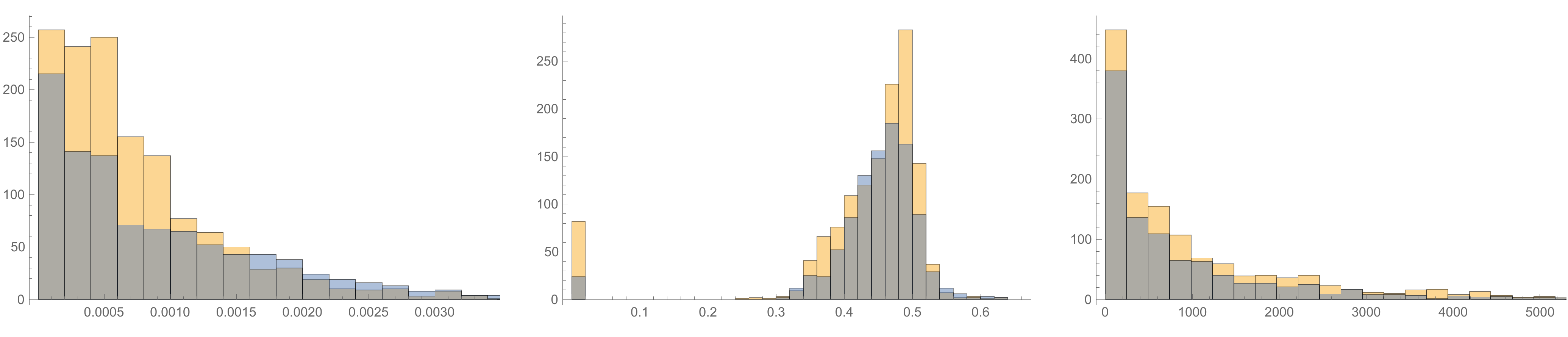}
    \put(3,-3){eigenvector centrality}
    \put(38,-3){closeness centrality}
     \put(71,-3){betweenness centrality}
    \end{overpic}
    \vspace{0.35cm}
    \caption{Histogram of the eigenvector, closeness, and betweenness centrality for the baseball and basketball-Twitter networks shown left, center, and right, respectively. Baseball is shown in orange and basketball is shown in blue in each histogram.}
    \label{fig:degreecentrality2}
\end{center}
\end{figure}

To give more evidence to the notion that the baseball Twitter and basketball Twitter networks have a similar structure, we note that the distribution of the networks' degrees (Figure \ref{fig:degreecentrality}), in-degrees, and out-degrees (Figure \ref{fig:degreecentrality1}) have very similar shapes and that the same holds for the networks' eigenvector, closeness, and betweenness centralities (Figure \ref{fig:degreecentrality2}). Here an individual's \textit{in-degree} is the number of Twitter followers they have while \textit{out-degree} is the number of player's they follow. An individual's \textit{eigenvector centrality} is high if they are followed by players that collectively have a high centrality. To have high \textit{closeness centrality} a player's mean distance to all other players in the network should be small. To have high \textit{betweenness centrality} the player should be on many of the shortest paths between other pairs of players. 

The top 5 players for each of these centralities for both the baseball-Twitter and basketball-Twitter networks are shown in Tables \ref{tab:basecentral} and \ref{tab:basketcentral}, respectively. Note that many of these players are the same suggesting that these are likely some of the most socially active members of the baseball and basketball community, at least as far as Twitter is concerned.    

\section{Conclusion}\label{sec:con}

In this paper we consider the question of ``Do social connections influence professionl group transition?'' in the context of both major league baseball and the National Basketball Association. Specifically, to what extent social connections can help predict how players change teams. We find that the addition of social data significantly improved the accuracy of our results.  In particular we compared which of the following types of data \textit{player performance}, \textit{team fitness}, and  \textit{social data} are more predictive in the context of machine learning.  We found that the addition of each of these data types does improve the predictive ability of our machine learning algorithms over random guessing.  In particular, knowing the social connections, and hence the social affinity scores of players significantly improves accuracy sometimes improving accuracy over 20\%.  

It is worth noting that, to our surprise, performative data and team fitness data made only small improvements to the accuracy of our machine learning algorithms.  There may be several reasons for this lack of improvement.  In separate experiments, we discovered that performative data does influence the likelihood of a player not returning to play the following year. That is, performance data seems to be better suited to answer ``if'' a player will leave a team rather than ``where'' the player will go. This is important in the sense that the number of players leaving the MLB and NBA is nearly equal to the number of players transitioning most years.  

In terms of team fitness, it may be possible to find a better indicator than the one we adopted since team fitness as defined had little effect on our prediction accuracy. It may also be the case that the lack of improvement in accuracy with the addition of this data may be due to the fact that healthy teams attract healthy players.  If this is the case considering a team's fitness would be equivalent to considering the fitness of an individual player.

We also note that the social networks under consideration are strikingly similar, hence the differences in prediction accuracy between baseball and basketball are likely not due to network structure. We conjecture that the differences in accuracy with the inclusion of social data between baseball and basketball may, in fact, be due to the percentage of players for which we have social data.  As further evidence we compare the accuracy of our machine learning algorithms on the early years of the data versus the later years of the data.  Both baseball and basketball show an increase in the percentage of players with social information and also an increase in the accuracy of the algorithm in the later years. College data, which we use as a proxy for social network information in the NBA, can also increase the the accuracy of our chosen algorithms, although how much depends on the algorithm used. We were able to obtain college information for a larger percentage of our basketball players and although the accuracy of the results did not improve as much as when we used Twitter data, the college data predates any transitions the players make in their professional careers while the Twitter data may not.

As mentioned, empirical data from early experiments show that performance data is a  good indicators of retirement.  Future work includes quantifying these results, and also investigating if a strong social network helps to delay retirement.  An interesting question to investigate is whether the inclusion of external social networks, for example between college and professional level coaches, would impact the results.  Finally, we wish to extend our results to other types of professional groups including groups that make up academic networks and industry networks to see if the impact of social networks is the same. 

\section{Appendix}
The appendix includes informative data that extends the data presented in the main body of the work.  It includes complete summary information for all of the machine learning algorithms utilized, and all of the combinations of features. It also includes the tables of the most socially active players in both baseball and basketball for all of the centralities we consider.

\begin{table}[h]
\begin{center}
\resizebox{\textwidth}{!}{
\begin{tabular}{|l|c|c|c|c|c|c|c|}
\hline
& Social & XGBoost & KNN & Extra & Random & ADA & Logistic\\
Features & Data & & & Trees & Forest & & Regression \\
\hline
Positions & N & 4.314\% & 4.358\% & 4.403\% & 4.955\% & 4.292\% & 3.827\%\\
Only & Y & 17.566\% & 5.685\% & 13.894\% & 16.172\% & 7.500\% & 13.496\%\\
\hline
Team & N & 4.270\% & 3.805\% & 4.380\% & 4.491\% & 3.761\% & 3.141\%\\
Only & Y & \textbf{19.955}\% & 5.088\% & 14.381\% & 17.5\% & 7.566\% & 19.912\%\\
\hline
Career & N & 4.579\% & 3.805\% & 5.221\% & 5.199\% & 4.513\% & 4.646\%\\
Length Only & Y & 17.278\% & 5.132\% & 14.004\% & 16.659 \% & 8.429\% & 13.849\%\\
\hline
Performance & N & 3.938\% & 4.890\% & 3.850\% & 3.805\% & 3.561\% & 4.314\%\\
Only & Y & 16.615\% & 5.353\% & 13.695\% & 15.376\% & 6.725\% & 13.207\%\\
\hline
Rank \& & N & 3.894\% & 3.872\% & 3.849\% & 3.894\% & 3.893\% & 4.004\%\\
Value Only & Y & 17.832\% & 5.088\% & 15.464\% & 16.946\% & 8.031\% & 14.491\%\\
\hline
Social Only & Y & 16.880\% & 5.376\% & 14.579\% & 16.372\% & 7.323\% & 15.332\%\\
\hline
All data & N & 4.358\% & 5.133\% & 4.159\% & 4.203\% & 4.270\% & 3.584\%\\
 & Y & 19.402\% & 5.508\% & 13.561\% & 16.460\% & 7.898\% & 18.075\%\\
 \hline
\end{tabular}
}
\end{center}
\caption{Summary of Algorithm Accuracy for Baseball data.  We observe that the highest accuracy was obtained using the XGBoost algorithm with the team and social data included. Using non-social features, the accuracy was about 1\% better than guessing, 3.44\%. Using social data, accuracy increased to 17-19\% in every case. }

\label{tab:baseball}
\end{table}

%

\begin{table}
\begin{center}
\resizebox{\textwidth}{!}{
\begin{tabular}{ |c|c|c|c|c|c|c|c|c|c| } 
\hline
Data Used & Twitter & College & XGBoost & KNN  & Extra & Random & ADA & Logistic\\
& & & & & Trees & Forest & & Regression\\
\hline
 & N & N & 3.506\% & 3.290\% & 3.506\% & 3.679\% & 3.549\% & 3.723\%\\
Positions & N & Y & 10.086\% & 13.810\% & 13.853\% & 13.853\% & 3.246\% & 4.155\%\\
Only  & Y & N & 26.190\% & 24.113\% & 26.277\% & 27.186\% & 78.788\% & 18.744\%\\
 & Y & Y & 27.143\% & 24.675\% & 28.095\% & 28.225\% & 5.757\% & 19.048\%\\
 
\hline
 & N & N &6.364\% & 4.761\% & 6.017\% & 5.714\% & 5.887\% & 5.757\%\\
Team & N & Y & 8.225\% & 11.125\% & 8.182\% & 8.831\% & 4.242\% & 5.411\% \\
Only & Y & N & 26.147\% & 23.636\% & 27.012\% & 28.225\% & 6.494\% & 14.675\%\\
 & Y & Y & 25.238\% & 23.810\% & 27.142\% & 26.580\% & 5.931\% & 14.372\%\\

\hline
 & N & N & 5.498\% & 4.026\% & 5.498\% & 5.802\% & 4.805\% & 5.628\%\\ 
Career & N & Y & 10.000\% & 19.523\% & 16.710\% & 16.410\% & 4.632\% & 6.234\% \\
Length Only & Y & N & 26.710\% & 23.766\% & 27.056\% & 26.536\% & 5.411\% & 19.004\% \\
 & Y & Y &26.190\% & 24.588\% & 27.099\% & 27.445\% & 5.974\% & 19.351\%\\ 

\hline 
 & N & N & 3.507\% & 3.420\% & 3.420\% & 3.377\% & 2.987\% & 3.593\%\\
Performance & N & Y & 5.324\% & 9.264\% & 6.926\% & 5.801\% & 4.242\% & 3.420\%\\
Only & Y & N & 21.904\% & 16.277\% & 26.493\% & 26.233\% & 6.017\% & 17.532\%\\
  & Y & Y & 22.597\% & 17.316\% & 28.009\% & 27.879\% & 3.593\% & 19.351\%\\
  
\hline
   & N & N & 5.833\% & 5.793\% & 5.952\% & 6.032\% & 3.948\% & 4.305\%\\
Rank \& & N & Y & 7.143\% & 5.456\% & 5.932\% & 5.793\% & 3.373\% & 4.484\%\\
Valuation & Y & N & 22.500\% & 6.111\% & 28.155\% & 27.301\% & 8.571\% & 18.214\% \\
 & Y & Y &24.782\% & 6.925\% & \textbf{30.238\%} & 29.642\% & 9.047\% & 19.246\%\\
\hline
Twitter & Y & N & 24.502\% & 22.208\% &26.104\% & 26.061\% & 6.580\% & 17.143\%\\ 

College & N & Y & 8.442\% & 9.913\% & 11.169\% & 10.563\% & 4.285\% & 4.156\% \\ 

All Social & Y & Y & 25.671\% & 23.982\% & 26.667\% & 26.623\% & 6.667\% & 19.004\%\\
\hline
 & N & N &7.056\% & 4.199\% & 8.138\% & 8.571\% & 3.853\%& 5.714\%\\ 
 All & N & Y & 8.009\% & 7.489\% & 10.390\% & 11.255\% & 4.156\% & 5.195\%\\
 Data & Y & N & 22.727\% & 5.844\% & 28.398\% & 27.402\% & 5.498\% & 14.372\%\\
 & Y & Y & 24.761\% & 19.524\% & 28.788\% & 29.740\% & 5.454\% & 14.805\%\\
\hline
\end{tabular}
}
\end{center}
\caption{Summary of algorithm accuracy for basketball data.    As with baseball, the inclusion of social data greatly increases the accuracy, sometimes by over 20\%.}
\label{tab:basketdata}
\end{table}

\begin{table}[h]
\begin{center}
\resizebox{\textwidth}{!}{
\begin{tabular}{|l|c|c|c|c|c|c|c|}
\hline
& College & XGBoost & KNN & Extra & Random & ADA & Logistic\\
Features & Data & & & Trees & Forest & & Regression \\
\hline
Positions & N & 4.110\% & 3.339\% & 4.057\% & 4.101\% & 3.888\% & 3.826\%\\
Only & Y & 10.738\% & 11.183\% & 11.379\% & 10.987\% & 3.683\% & 3.692\%\\
\hline
Team & N & 4.430\% & 4.110\% & 4.404\% & 4.377\% & 4.164\% & 4.093\%\\
Only & Y & 8.745\% & 10.338\% & 8.585\% & 8.532\% & 4.350\% & 3.986\%\\
\hline
Career & N & 4.893\% & 3.817\% & 4.893\% & 4.706\% & 4.866\% & 4.715\%\\
Length Only & Y & 14.430\% & \textbf{16.272}\% & 14.617\% & 14.920\% & 3.995\% & 4.662\%\\
\hline
Performance & N & 3.549\% & 3.7010\% & 3.710\% & 3.647\% & 3.816\% & 3.932\%\\
Only & Y & 4.804\% & 6.076\% & 6.032\% & 5.133\% & 3.523\% & 4.235\%\\
\hline
Rank \& & N & 6.761\% & 6.236\% & 6.183\% & 5.943\% & 4.226\% & 3.727\%\\
Value Only & Y & 7.500\% & 5.489\% & 5.267\% & 5.872\% & 4.448\% & 3.665\%\\
\hline
College Only & Y & 8.398\% & 10.035\% & 8.959\% & 8.816\% & 3.879\% & 3.994\%\\
\hline
All data & N & 6.014\% & 5.382\% & 7.117\% & 6.993\% & 4.190\% & 4.350\%\\
 & Y & 7.256\% & 6.316\% & 8.434\% & 8.825\% & 4.377\% & 4.092\%\\
 \hline
\end{tabular}
}
\end{center}
\caption{Summary of Algorithm Accuracy for Basketball College data.  We observe that the highest accuracy was obtained using the XGBoost algorithm with the team and social data included. Using non-social features, the accuracy was about 1\% better than guessing, 3.44\%. Using social data, accuracy increased to 17-19\% in every case. }
\label{tab:collegedata}
\end{table}

\begin{table*}
\centering
\resizebox{\textwidth}{!}{
\begin{tabular}{c  c  c  c  c  c }
\hline
\hline
Rank      &  1               & 2               & 3               & 4            & 5         \\
\hline
DegC       &  Jose Bautista   & Ervin Santana   & Frank Garces    & Rob Wooten   & Mike Trout\\      
EigC       &  Mike Trout      & David Price     & Jose Bautista   & Kevin Millar & Torii Hunter\\  
CloseC     &  Jose Bautista   & Frank Garces    & Ervin Santana   & Rob Wooten   & Brian Bannister\\  
BetC       &  Jose Bautista   & Ervin Santana   & Mike Trout      & David Price  & Will Middlebrooks\\ 
\hline
\hline
\end{tabular}
}
\vspace{.1cm}
\caption{The top 5 baseball players of the baseball-Twitter network based on degree centrality (DegC), eigenvector centrality (EigC), closeness centrality (CloseC), and betweenness centrality (BetC).}
\label{tab:basecentral}
\end{table*}

\begin{table*}
\centering
\resizebox{\textwidth}{!}{
\begin{tabular}{c  c  c  c  c  c }
\hline
\hline
Rank      &  1               & 2               & 3               & 4            & 5         \\
\hline
DegC       &  Jamal Crawford & Baron Davis     & Joe Smith       & Kevin Durant & John Wall\\      
EigC       &  Kevin Durant   & LeBron James    & Jamal Crawford  & Kobe Bryant  & Baron Davis\\  
CloseC     &  Joe Smith      & Maurice Ager    & Tracy Murray    & Jamal Crawford & Baron Davis\\  
BetC       &  Jamal Crawford & Baron Davis     & John Wall       & Kevin Durant   & Maurice Ager\\ 
\hline
\hline
\end{tabular}
}
\vspace{.1cm}
\caption{The top 5 basketball players of the basketball-Twitter network based on degree centrality (DegC), eigenvector centrality (EigC), closeness centrality (CloseC), and betweenness centrality (BetC).}
\label{tab:basketcentral}
\end{table*}

\FloatBarrier
\section*{References}
\bibliography{references}{}

\end{document}